\documentclass[aps,twocolumn,groupedaddress,nofootinbib]{revtex4-1}
\usepackage{amssymb, amsmath, siunitx}
\usepackage{hyperref}
\usepackage{slashed}
\usepackage{bbold}
\usepackage{graphicx}

\begin{document}


\title{Starobinsky-type Inflation in Dynamical Supergravity Breaking Scenarios}

\author{Jean Alexandre$^a$, Nick Houston$^a$ and Nick E. Mavromatos$^{a,b}$}

\affiliation{$^a$Theoretical Particle Physics and Cosmology Group, Physics Department, King's College London, Strand, London WC2R 2LS. \\
 $^b$ Also currently at:  Theory Division, Physics Department, CERN, CH-1211 Geneva 23, Switzerland.}


\begin{flushleft}
KCL-PH-TH/2013-\textbf{39} \\
LCTS/2013-26
\end{flushleft} 

\begin{abstract} 
\vspace{0.2cm}
In the context of dynamical breaking of local supersymmetry (supergravity), including the Deser-Zumino super-Higgs effect, for the simple but quite representative cases of $\mathcal{N}=1$, $D=4$ supergravity, we discuss the emergence of Starobinsky-type inflation, due 
to quantum corrections in the effective action arising from integrating out gravitino fields in their massive phase. This type of inflation may occur after a first-stage small-field inflation that characterises models near the origin of the one-loop effective potential, and it may occur at the non-trivial minima of the latter. Phenomenologically realistic scenarios, compatible with the Planck data, may be expected for the conformal supergravity variants of the basic model.
\end{abstract}

\maketitle

This short article serves as an \emph{addendum} to our previous publication~\cite{ahm}, where we discussed dynamical breaking of 
supergravity (SUGRA) theories via gravitino condensation. In particular, we shall demonstrate the compatibility of this scenario with 
Starobinsky-like~\cite{staro} inflationary scenarios, which in our case can characterise the massive gravitino phase. As we shall argue, this is a second
inflationary phase, that may succeed a first inflation which occurs in the flat region of the one-loop effective potential for the gravitino condensate field~\cite{emdyno}. 

Starobinsky inflation is a model for obtaining a de Sitter (inflationary) cosmological solution to gravitational equations arising from a
(four space-time-dimensional) action that includes higher curvature terms, specifically of the type in which the quadratic curvature corrections consist only of scalar curvature terms~\cite{staro}
\begin{eqnarray}\label{staroaction}
{\mathcal S} = \frac{1}{2 \, \kappa^2 } \, \int d^4 x \sqrt{-g}\,  \left(R  + \beta  \, R^2 \right) ~,~ 
\beta = \frac{8\, \pi}{3\, {\mathcal M}^2 }~,
\end{eqnarray}
where $\kappa^2=8\pi G$, and ${\rm G}=1/m_P^2$ is Newton's (gravitational) constant in four space-time dimensions, with $m_P$ the Planck mass, and ${\mathcal M}$ is a constant of mass dimension one, characteristic of the model. 

The important feature of this model is that inflationary dynamics are driven by the purely gravitational sector, through the $R^2$ terms, 
and the scale of inflation is linked to ${\mathcal M}$. From a microscopic point of view, the scalar curvature-squared terms in (\ref{staroaction}) are viewed as the result of \emph{quantum fluctuations} (at one-loop level)  of conformal (massless or high energy) \emph{matter fields} of various spins, which have been integrated out in the relevant path integral in a curved background space-time~\cite{loop}. The quantum mechanics of this model, by means of \emph{tunneling} of the Universe from a state of ``nothing'' to the inflationary phase of ref.~\cite{staro} has been discussed in detail in \cite{vilenkin}.
The above considerations necessitate truncation to one-loop quantum order and to curvature-square (four-derivative) terms, which 
implies that there must be a region of validity for curvature invariants such that $\mathcal{O}\big(R^2/m_p^4\big) \ll 1$, which is a condition satisfied in phenomenologically realistic scenarios of inflation~\cite{encyclo,Planck}, for which the inflationary Hubble scale $H_I \leq  0.74 \times 10^{-5} \, m_P = {\mathcal O}(10^{15})~{\rm GeV}$ (the reader should recall that $R \propto H_I^2$ in the inflationary phase). 

\begin{figure}[h!!!]
\centering
		\includegraphics[width=0.45\textwidth]{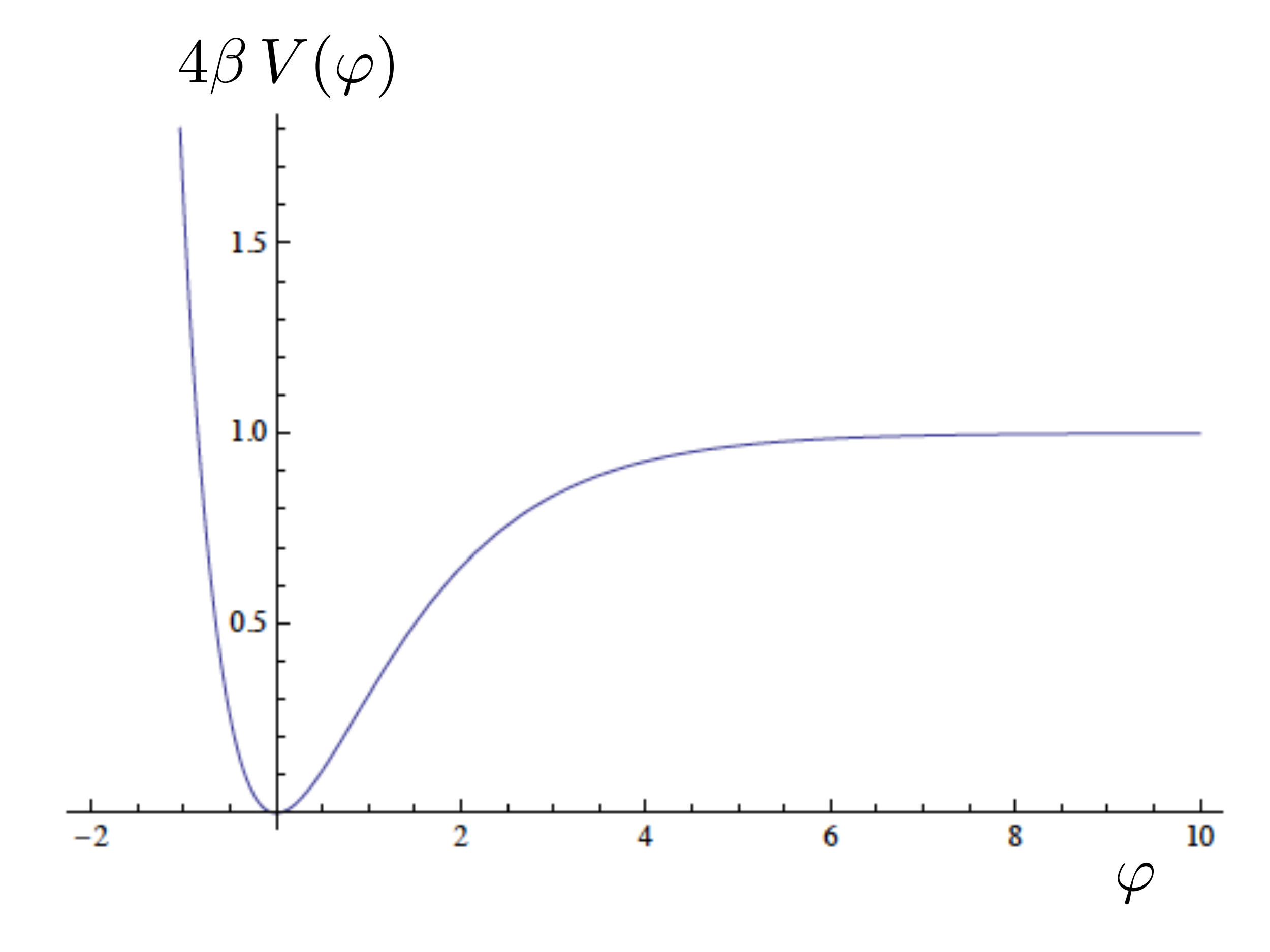}
		\caption{The effective potential (\ref{staropotent}) of the collective scalar field $\varphi$ that describes the one-loop quantum fluctuations of matter fields, leading to the higher-order scalar curvature corrections in the Starobinski model for inflation (\ref{staroaction}). The potential is sufficiently flat to ensure slow-roll conditions for inflation are satisfied, in agreement with the Planck data, for appropriate values of the scale $1/\beta \propto {\mathcal M}^2$ (which sets the overall scale of inflation in the model).}
\label{fig:potstar}	
\end{figure}

Although the inflation in this model is not driven by fundamental rolling scalar fields, nevertheless the model (\ref{staroaction}) (and for that matter, any other model where the Einstein-Hilbert space-time Lagrangian density is replaced by an arbitrary function $f(R)$ of the scalar curvature)  is \emph{conformally equivalent} to that of an ordinary Einstein-gravity coupled to a scalar field with a potential that drives inflation~\cite{whitt}. To see this, one firstly linearises the $R^2$ terms in (\ref{staroaction}) by means of an auxiliary (Lagrange-multiplier) field $\tilde \alpha (x)$, before rescaling the metric by a conformal transformation and redefining the scalar field 
(so that the final theory acquires canonically-normalised Einstein and scalar-field terms):
\begin{eqnarray}\label{confmetric}
&&g_{\mu\nu} \rightarrow g^E_{\mu\nu} = \left(1 + 2 \, \beta \, {\tilde \alpha (x)} \right) \, g_{\mu\nu} ~, \\
&& \tilde \alpha \left(x\right) \to \varphi (x) \equiv \sqrt{\frac{3}{2}} \, {\rm ln} \, \left(1 + 2\, \beta \, {\tilde \alpha \left(x\right)} \right)~.
\end{eqnarray}
These steps may be understood schematically via
\begin{align}\label{steps}
	&\int d^4 x \sqrt{-g}\,  \left( R  + \beta  \, R^2 \right)  \\
  	&\hookrightarrow\int d^4 x \sqrt{-g}\,  \left(  \left(1 + 2\, \beta \, \tilde \alpha \left(x\right) \right) \, R  -  \beta  \, {\tilde \alpha (x)}^2 \right)\nonumber  \\
    &\hookrightarrow\int d^4 x \sqrt{-g^E}\,  \left(R^E +  g^{E\, \mu\, \nu} \, \partial_\mu \, \varphi \, \partial_\nu \, \varphi - V\right(\varphi\left) \right)~,\nonumber
\end{align}
where the arrows have the meaning that the corresponding actions appear in the appropriate path integrals, 
with the potential $V(\varphi)$ given by:
\begin{eqnarray}\label{staropotent}
 V(\varphi ) = \frac{\left( 1 - e^{-\sqrt{\frac{2}{3}} \, \varphi } \right)^2}{4\, \beta} \, 
  = \frac{3 {\mathcal M}^2 \, \Big( 1 - e^{-\sqrt{\frac{2}{3}} \, \varphi } \Big)^2}{32\, \pi }  \,  ~.
\end{eqnarray}
The potential is plotted in fig.~\ref{fig:potstar}. 
We observe that it is sufficiently flat for large values of $\varphi$ (compared to the Planck scale) to produce phenomenologically acceptable inflation, with the scalar field $\varphi$ playing the role of the inflaton. 
In fact the Starobinsky model fits excellently the Planck data on inflation~\cite{Planck}. 

Quantum-gravity corrections in the original Starobinsky model (\ref{staroaction}) have been considered recently in \cite{copeland} from the point of view of an \emph{exact renormalisation-group} (RG) analysis~\cite{litim}. 
It was shown that the \emph{non-perturbative} beta-functions for the `running' of Newton's `constant' G and the dimensionless $R^2$ coupling $\beta^{-1}$ in (\ref{staroaction}) imply an \emph{asymptotically safe} Ultraviolet (UV) fixed point 
for the former (that is, G($k \to \infty$) $ \to $ constant, for some 4-momentum cutoff scale $k$)), in the spirit of Weinberg~\cite{weinberg}, and an
attractive \emph{asymptotically-free} ($\beta^{-1} (k \to \infty) \to 0$)
point for the latter. 
In this sense, the smallness of the $R^2$ coupling, required for agreement with
inflationary observables~\cite{Planck}, is naturally ensured by the presence of the asymptotically free UV fixed
point. 

The agreement of the model of \cite{staro} with the Planck data triggered an enormous interest in the current literature
in revisiting the model from various points of view, such as its connection with no-scale supergravity~\cite{olive} and (super)conformal versions of supergravity and related areas~\cite{sugrainfl}. 
In the latter works however the Starobinski scalar field is fundamental, arising from the appropriate scalar component of some chiral superfield that appears in the superpotentials of the model. 
Although of great value, illuminating a strong connection between supergravity models and inflationary physics, and especially for explaining the low-scale of inflation compared to the Planck scale, these works contradict the original spirit of the Starobinsky model (\ref{staroaction}) where, as mentioned previously, the higher curvature corrections are viewed as arising from quantum fluctuations of matter fields in a curved space-time background such that inflation is driven by the pure gravity sector in the absence of fundamental scalars. 

In a recent publication~\cite{emdyno}, we have considered an alternative inflationary scenario, in which, in the spirit of the original Starobinsky  model, the inflaton field was not a fundamental scalar but arose as a result of condensation (in the scalar $s$-wave channel) of the gravitino field in simple supergravity (SUGRA) models with spontaneous breaking of global supersymmetry (SUSY) via the super-Higgs effect~\cite{ahm}, at a (mass) scale $\sqrt{f}$. 
Dynamical breaking of SUGRA, in the sense of the generation of a mass for the gravitino field $\psi_\mu$, whilst the gravitons remain massless, occurs in the model as a result of the four-gravitino interactions characterising the SUGRA action, arising from the torsionful contributions of the spin connection, characteristic of local supersymmetric theories. 
The one-loop effective potential for the scalar gravitino condensate field $\sigma_c \propto \langle \overline \psi_\mu \,  \psi^\mu \rangle $ has a double-well shape as a function of $\sigma_c$ which is symmetric about the origin, as dictated by the fact that the sign of a fermion mass does not have physical significance.  
Dynamical generation of the gravitino mass occurs at the non-trivial minima corresponding to $\sigma_c  \ne 0$.  
The potential of the $\sigma_c$ field is also flat near the origin, and this has been identified in \cite{emdyno} with the inflationary phase.

In \cite{ahm} the one-loop effective potential was derived by first formulating the theory on a curved de Sitter background~\cite{fradkin}, with cosmological constant (one-loop induced) $\Lambda > 0$, and integrating out spin-2 (graviton) and spin 3/2 (gravitino)
quantum fluctuations in a given class of gauges (\emph{physical}), before considering the flat limit $\Lambda \to 0$ in a self-consistent way. The detailed analysis in \cite{ahm}, performed in the \emph{physical gauge}, has demonstrated that the dynamically broken phase is then stable (in the sense of the effective action not being characterised by imaginary parts) provided 
\begin{equation}\label{absimparts}
\sigma_c^2 \le f^2 ~.
\end{equation}
This result demonstrated the importance of the existence of global SUSY breaking scale for the stability of the phase where dynamical generation of gravitino masses occurs, which was not considered in the previous literature~\cite{odintsov}~\footnote{Although performed 
in different gauges to our own, the result of those references that imaginary parts prevent gravitino mass generation would also be valid in the case we consider here, were it not for the super-Higgs effect and the condition (\ref{absimparts}). Such a conclusion could however not be reached in \cite{odintsov}, as the role of the super-Higgs effect, and the `eating' of the Goldstino associated with the global SUSY breaking by the gravitino, was not included.}. 

The self-consistency of the $\Lambda \to 0$ limit necessarily implies the vanishing of the one-loop effective potential at the non-trivial minima, which is a limiting case consistent with the supersymmetry breaking. This restricts the scale of the $f^2$ and $\sigma_c^2$ in such a way that both scales must be of order of Planck if the simplest four dimensional  ${\mathcal N}=1 $ SUGRA model is considered.
On the other hand, if one considers superconformal versions of SUGRA, \emph{e.g}. those in ref.~\cite{confsugra}, then phenomenologically realistic scales for $f^2$ and $\sigma_c$ of order of the Grand Unification Scale, can appear,
for appropriate values of the expectation value of the conformal factor, implying inflationary scenarios in perfect agreement with the Planck data~\cite{emdyno,Planck}, on equal footing to the original Starobinski model. 
The inflationary period in this scenario is obtained by a simple embedding of the one-loop effective potential for the gravitino condensate field in a standard Einstein-background gravity, where higher curvature corrections are ignored, whilst 
the end of the inflationary period  coincides with the flat space-time limit that characterises the dynamical breaking of SUGRA at the non-trivial minima of the one-loop effective potential.

In this note we would like to consider an extension of the analysis of \cite{ahm}, where the de Sitter parameter $\Lambda$ is perturbatively small compared to $m_P^2$, but not zero, so that truncation of the series to order $\Lambda^2$ suffices. 
This is in the spirit of the original Starobinsky model~\cite{staro}, with the role of matter fulfilled by the now-massive gravitino field.
Specifically, we are interested in the behaviour of the effective potential near the non-trivial minimum, where $\sigma_c $ is a non-zero constant. 
In our analysis, unlike Starobinsky's original work, we will keep the contributions from \emph{both} graviton (spin-two) and gravitino quantum fluctuations.
Notice that our one-loop analysis does not allow us to make any comment on asymptotic safety of the solution as in \cite{copeland}, as this would require detailed analysis based on exact RG which we do not perform here. 

We firstly note that the one-loop effective potential, obtained by integrating out gravitons and (massive) gravitino fields in the scalar channel (after appropriate euclideanisation), may be expressed as a power series in $\Lambda$: 
\begin{align}\label{effactionl2}
	\Gamma\simeq S_{\rm cl}-\frac{24\pi^2}{\Lambda^2 }\big(&\alpha^F_0+\alpha_0^B
	+ \left(\alpha^F_{1}+ \alpha^B_{1}\right)\Lambda\nonumber\\
	&\qquad+\left(\alpha^F_{2}+ \alpha^B_{2}\right)\Lambda^2+\dots\big)~,
\end{align} where $S_{\rm cl}$ denotes the classical action with tree-level cosmological constant $\Lambda_0$ (to be contrasted with the one-loop cosmological constant $\Lambda$):
	\begin{align}
		-\frac{1}{2\kappa^2}\int d^4 x \sqrt{g}\left(\widehat{R}-2\Lambda_0\right), \quad
		\Lambda_0=\kappa^2\left(\sigma^2-f^2\right)~,
	\end{align} 
with $\widehat R$ denoting the fixed $S^4$ background we expand around ($\widehat R=4\Lambda$, Volume = $24\pi^2/\Lambda^2$), and the $\alpha$'s indicate the bosonic (graviton) and fermionic (gravitino) quantum corrections at each order in $\Lambda$.

The leading order term in $\Lambda$ is then the effective action found in \cite{ahm} in the limit $\Lambda\to0$, 
	\begin{align}
		\Gamma_{\Lambda\to0}\simeq-\frac{24\pi^2}{\Lambda^2}\left(-\frac{\Lambda_0}{\kappa^2}+\alpha_0^F+\alpha_0^B\right)
		\equiv\frac{24\pi^2}{\Lambda^2}\frac{\Lambda_1}{\kappa^2},
	\end{align}
and the remaining quantum corrections then, proportional to $\Lambda$ and $\Lambda^2$ may be identified respectively with Einstein-Hilbert $R$-type and Starobinsky $R^2$-type terms in an effective action (\ref{effactionl3}) of the form
\begin{align}\label{effactionl3}
\Gamma\simeq&-\frac{1}{2\kappa^2} \int d^4 x \sqrt{g} \left(\left(\widehat R-2\Lambda_1\right)  +\alpha_1 \, \widehat R+ \alpha_2 \, \widehat R^2\right)~,
\end{align}
where we have combined terms of order $\Lambda^2$ into curvature scalar square terms. For general backgrounds such terms 
would correspond to invariants of the form ${\widehat R}_{\mu\nu\rho\sigma} \, {\widehat R}^{\mu\nu\rho\sigma} $, ${\widehat R}_{\mu\nu} \, {\widehat R}^{\mu\nu}$ and ${\widehat R}^2$, which for a de Sitter background all combine to yield ${\widehat R}^2$ terms. 
The coefficients $\alpha_1$  and $\alpha_2$ absorb the non-polynomial (logarithmic) in $\Lambda$ contributions, so that we may then identify \eqref{effactionl3} with \eqref{effactionl2} via 
	\begin{align}\label{alpha}
		\alpha_1=\frac{\kappa^2}{2}\left(\alpha^F_1+\alpha^B_1\right)~,\quad
		\alpha_2=\frac{\kappa^2}{8}\left(\alpha^F_2+\alpha^B_2\right)~.
	\end{align}

To identify the conditions for phenomenologically acceptable Starobinsky inflation around the non-trivial minima of the broken SUGRA phase 
of our model, we impose first the \emph{cancelation} of the ``classical'' Einstein-Hilbert space term $\widehat R $ by the ``cosmological constant'' term $\Lambda_1$, i.e. that $\widehat R = 4 \, \Lambda = 2\, \Lambda_1 $.
This condition should be understood as a necessary one characterising our background in order to produce phenomenologically-acceptable 
Starobinsky inflation in the broken SUGRA phase following the first inflationary stage, as discussed in \cite{emdyno}. 
This may naturally be understood as a generalisation of the relation $\widehat R=2\Lambda_1=0$, imposed in \cite{ahm} as a self-consistency condition for the dynamical generation of a gravitino mass.

The effective Newton's constant in  (\ref{effactionl3}) is then $\kappa_{\rm eff}^2=\kappa^2/\alpha_1$, and from this, we can express the effective Starobinsky scale (\ref{staroaction}) in terms of $\kappa_{\rm eff}$ as $\beta_{\rm eff} \equiv  \alpha_2/\alpha_1$.
This condition thus makes a direct link between the action (\ref{effactionl2}) with a Starobinsky type action (\ref{staroaction}).
Comparing with (\ref{staroaction}), we may then determine the Starobinsky inflationary scale in this case as
\begin{equation}\label{staroours}
{\mathcal M} = \sqrt{\frac{8 \pi}{3} \, \frac{\alpha_1}{\alpha_2} }~.
\end{equation}

We may then determine the coefficients $\alpha_1$ and $\alpha_2$ in order to evaluate the scale $1/\beta$ of the effective Starobinsky potential given in fig.~\ref{fig:potstar} in this case, and thus the scale of the second inflationary phase. 
To this end, we use the results of \cite{ahm}, derived via an asymptotic expansion as explained in the appendix therein, to obtain the following forms for the coefficients 			
	\begin{eqnarray}\label{aif}
		\alpha^F_1&=\frac{\left(25491-5 \sqrt{27076337}\right)}{25016} \tilde\kappa^2 \sigma_c ^2 \log \left(\frac{\Lambda}{\mu^2}\right) \nonumber \\
		&+\frac{\left(3 \sqrt{65028102}-18700\right) }{81397}\tilde\kappa^2 \sigma_c ^2 \nonumber \\
		&+\frac{\left(\sqrt{100304662585}-247787\right) }{945888}\tilde\kappa^2 \sigma_c ^2 \log \left(\frac{\tilde\kappa^2\sigma_c^2}{\mu^2} \right)~, \nonumber \\
		\alpha^F_{2}&=\frac{\left(6 \sqrt{5018206}-12882\right)}{38914}\log \left(\frac{\tilde\kappa^2\sigma_c^2}{\mu^2}\right) \nonumber \\
		&+\frac{\left(50249-\sqrt{2590498021}\right) }{22066}\log \left(\frac{\Lambda}{\mu^2}\right) \nonumber \\
		&+\frac{\sqrt{10592733}-1377}{65388}~,
	\end{eqnarray}
and 
	\begin{eqnarray}\label{aib}
		\alpha^B_1&=
		\frac{\sqrt{356979979}-17707}{64839}\Lambda_0 \log \left(\frac{\Lambda }{3 \mu ^2}\right) \nonumber \\
		&+\frac{\left(\sqrt{2812791101}-52583\right) }{9244}\Lambda_0 \log \left(-\frac{3 \Lambda_0}{\mu ^2}\right)
		\nonumber \\
		&-\frac{\left(\sqrt{1416210349}-27907\right)\left(1+\log\left(2\right)\right)}{198570} \Lambda_0~, \nonumber \\
		\alpha^B_{2}&=-\frac{\left(\sqrt{220573721}-19811\right) }{232300}\log \left(\frac{\Lambda }{3 \mu ^2}\right)
		\nonumber \\
		&+\frac{\left(10 \sqrt{12614479}-36763\right) }{86027}\log \left(-\frac{6 \Lambda_0}{\mu ^2}\right) \nonumber \\
		&+\frac{2731-\sqrt{1392978}}{76777}~,
	\end{eqnarray}
where ${\tilde \kappa} = e^{-\langle \Phi  \rangle } \, \kappa$ is the conformally-rescaled gravitational constant in the model of \cite{confsugra} and $\langle  \Phi \rangle \ne 0$ is the v.e.v. of the conformal (`dilaton') factor, assumed to be stabilised by means of an appropriate potential. In the case of standard ${\mathcal N}=1$ SUGRA, $\langle \Phi \rangle = 0$.  We note at this stage that the spin-two parts, arising from integrating out graviton quantum fluctuations, are not dominant in the conformal case~\cite{ahm}, provided ${\tilde \kappa}/\kappa \ge {\mathcal O}(10^3)$, which leads~\cite{emdyno} to the agreement of the first  inflationary phase of the model with the Planck data~\cite{Planck}.  
However, if the first phase is succeeded by a Starobinsky phase, it is the latter only that needs to be checked against the data. 

We search numerically for points in the parameter space such that; the effective equations
	\begin{align}
		 \frac{\partial\Gamma}{\partial\Lambda}=0~, \quad
		 \frac{\partial\Gamma}{\partial\sigma}=0~,
	\end{align}
are satisfied, $\Lambda$ is small and positive ($0<\Lambda<10^{-5}M^2_{\rm Pl}$, to ensure the validity of our expansion in $\Lambda$) and $10^{-6}<\mathcal{M}/M_{\rm Pl}<10^{-4}$, to match with known phenomenology of  Starobinsky inflation \cite{Planck}. 

\begin{figure}[h!!!]
	\centering
		\includegraphics[width=0.45\textwidth]{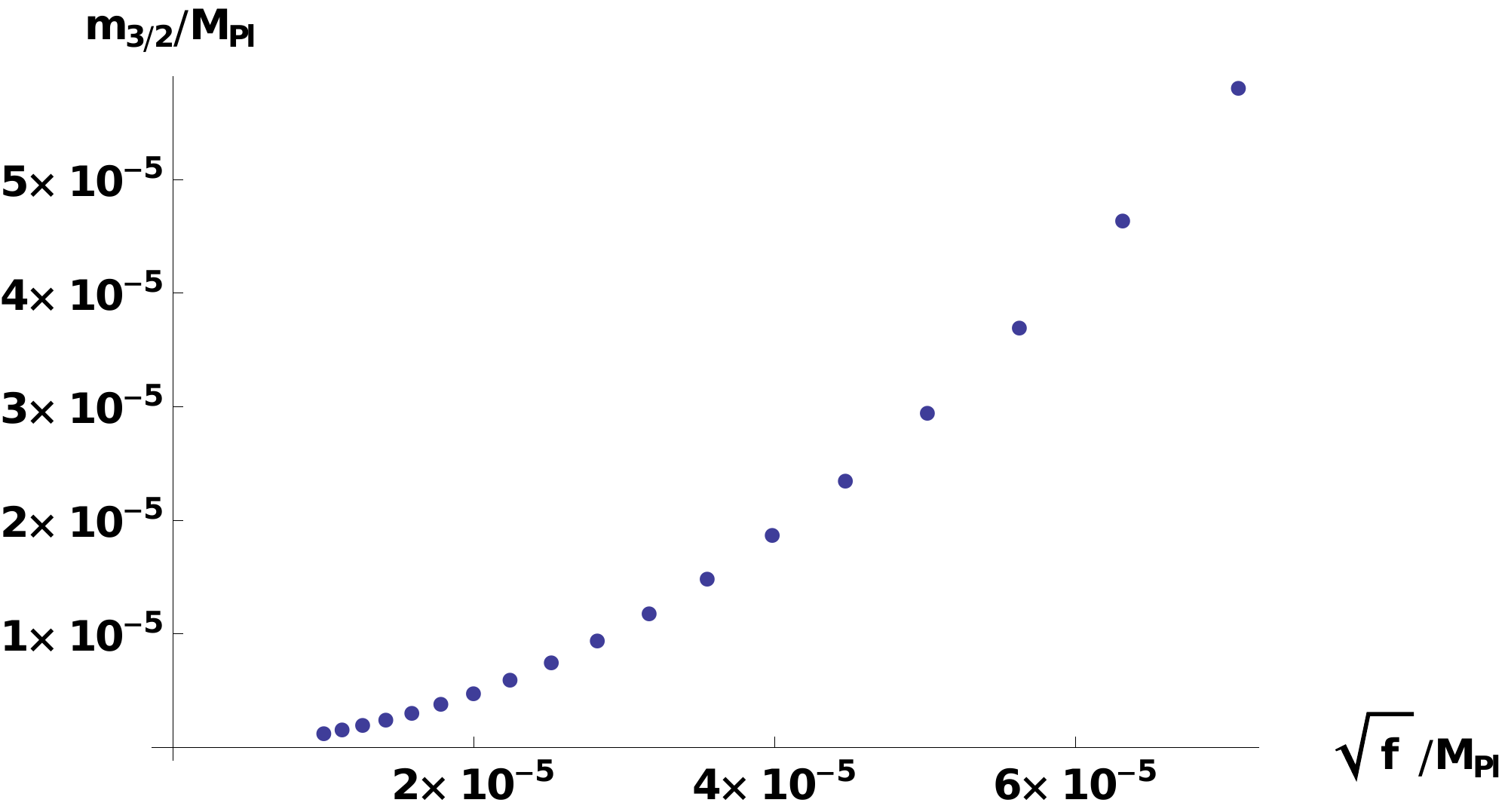}
		\caption{Results for $\tilde \kappa=10^3\kappa$.}
		\label{fig:2a}
		\end{figure}

\begin{figure}[h!!!]
	\centering		
	\includegraphics[width=0.45\textwidth]{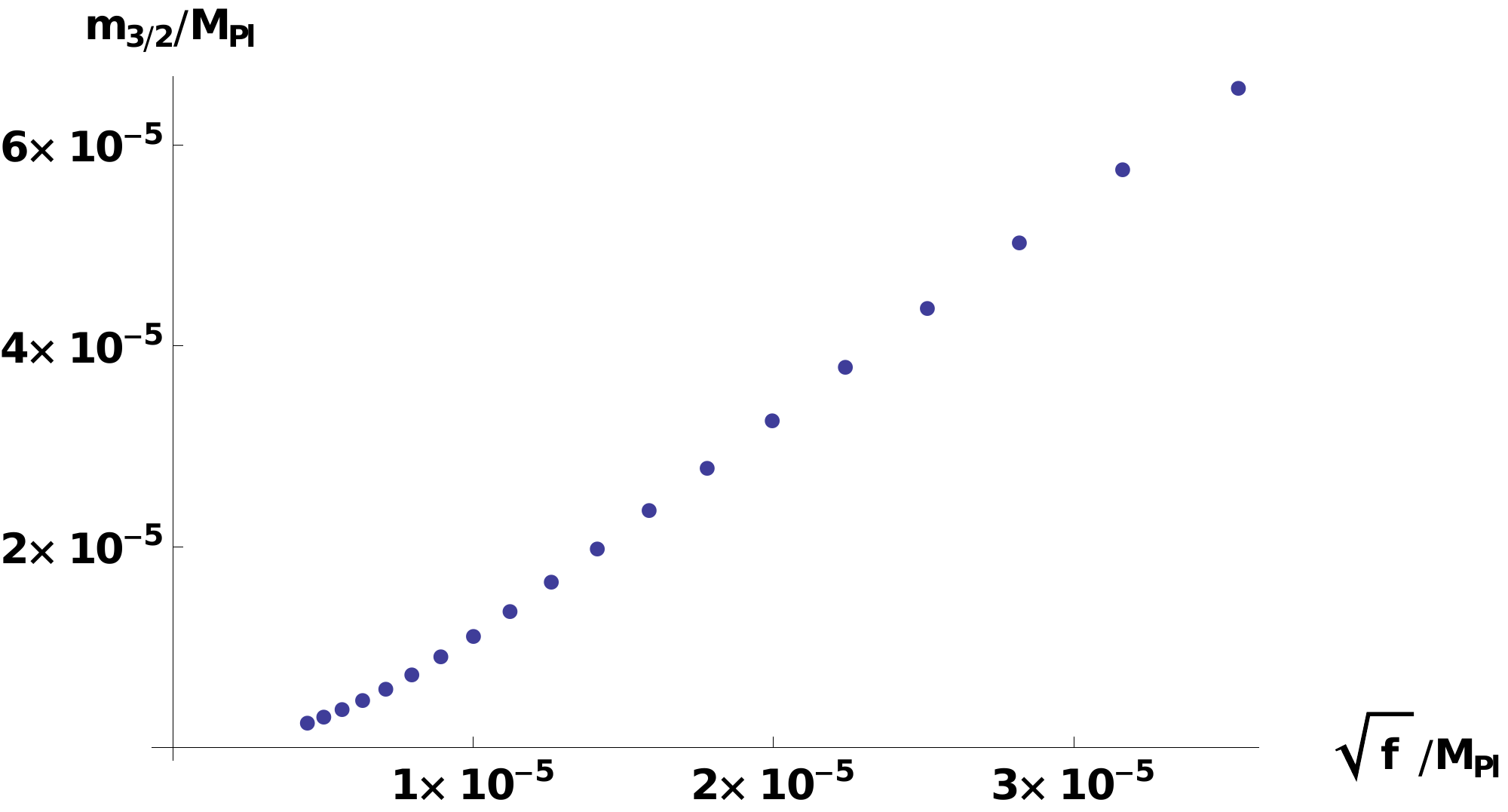}
		\caption{Results for $\tilde \kappa=10^4\kappa$.}
		\label{fig:2b}
	\end{figure}

For $\tilde \kappa=\kappa$ (i.e. for non-conformal supergravity), we were unable to find any solutions satisfying these constraints. 
This of course may not be surprising, given the previously demonstrated non-phenomenological suitability of this simple model \cite{ahm}. 
If we consider $\tilde \kappa>>\kappa$ however, we find that we are able to satisfy the above constraints for a range of values. 
We present this via the two representative cases below, indicated in figs.~\ref{fig:2a}, \ref{fig:2b}, where we have the gravitino mass \cite{ahm}
	\begin{align}
		m_{3/2}=\sqrt\frac{11}{2}\tilde\kappa\sigma_c~,
	\end{align}
$\sqrt{f}$ is the scale of global supersymmetry breaking, and we have set the normalisation scale via $\kappa\mu=\sqrt{8\pi}$.\\
Every point in the graphs of the figures is selected to make the Starobinsky scale of order ${\mathcal M} \sim 10^{-5} \, M_{\rm Pl} $, hence we able to achieve phenomenologically acceptable Starobinsky inflation in the massive gravitino phase, consistent with the Planck-satellite data~\cite{Planck}. 

Exit from the inflationary phase is a complicated issue which we shall not discuss here, aside from the observation that it can be achieved by
coherent oscillations of the gravitino condensate field around its minima, or tunnelling processes \`a  la Vilenkin~\cite{vilenkin}. We hope to address these issues in detail in a future work.

\section*{Acknowledgements}

The work of N.H. is supported by a KCL GTA studentship, while that of 
N.E.M. is supported in part by the London Centre for Terauniverse Studies (LCTS), using funding from the European Research Council via the Advanced Investigator Grant 267352 and by STFC (UK) under the research grant ST/J002798/1.

  \end{document}